\documentclass[fleqn,12pt,twoside]{article}
\usepackage[headings]{espcrc1}

\readRCS $Id: espcrc1.tex,v 1.2 2004/02/24 11:22:11 spepping Exp $
\usepackage{graphicx}
\usepackage[figuresright]{rotating}

\newcommand{\AmS}{{\protect\the\textfont2
  A\kern-.1667em\lower.5ex\hbox{M}\kern-.125emS}}

\title{In-medium Hadrons -- Properties, Interaction and Formation}
\author{Kai Gallmeister\addressmark[UGi]\address{Institut
fuer Theoretische Physik, Universitaet Giessen, D-35392 Giessen,
Germany}, Tina Leitner\addressmark, Stefan Leupold\addressmark,
Ulrich Mosel\addressmark\thanks{e-mail:
mosel@physik.uni-giessen.de},Pascal Muehlich\addressmark, Luis
Alvarez-Ruso\addressmark, Vitaly Shklyar\addressmark\thanks{Work
supported by BMBF and DFG}}

\runtitle{In-medium Hadrons} \runauthor{U. Mosel}

\begin{document}

\maketitle

\begin{abstract}
In this talk various aspects of in-medium behavior of hadrons are
discussed with an emphasis on observable effects. Examples for
theoretical predictions of in-medium spectral functions are given
and the importance of resonance-hole excitations is stressed. It is
also stressed that final state interactions can have a major effect
on observables and thus have to be considered as part of the theory.
This is demonstrated with examples from neutrino-nucleus
interactions. Finally, the possibility to access hadron formation
times in high-energy photonuclear (or neutrino-induced) reactions is
illustrated.
\end{abstract}

\section{Introduction}
Hadrons, embedded inside nuclei, obviously change some of their
properties. They acquire complex selfenergies with the real parts
reflecting the binding (or non-binding) properties and the imaginary
parts reflecting the interactions and possibly their changes inside
the medium. Particles that are produced through resonances or -- at
high energies -- through strings become physical, on-shell particles
only after some formation time. In this case the nuclear medium may
affect the formation process and can thus act as a  micro-detector
for the early stages of particle production.

Naively, one expects that in lowest order all in-medium effects go
linearly with the density of nuclear matter, $\rho$, around the
hadron. This has triggered a series of experiments with relativistic
and ultrarelativistic heavy-ions, which can reach high densities,
that have looked for such effects and have indeed reported in-medium
changes of the $\rho$ meson \cite{Ceres,Xu,NA60}. However, it has
been pointed out quite early \cite{Mosel} that also experiments with
microscopic probes on nuclei can yield in-medium signals that are as
large as those obtained in heavy-ion collisions. Although, of
course, the density probed here is always below $\rho_0$ the
observed signal is cleaner in the sense that it does not contain an
implicit integration over very different phases of the reaction and
the nuclear environment. The signal to be expected is also nearly as
large as that seen in heavy-ion collisions. This idea has been
followed up in recent experiments with photons on nuclei
\cite{g7,Trnka}, where indeed changes of the $\omega$ meson in
medium have been reported \cite{Trnka}.

In this talk I will give an overview of this field with an emphasis
on observable effects. I will also stress the need for a theoretical
framework that allows to follow the signals from the hadron inside
nuclei all the way to the final detector. More details can be found
in two previous reviews \cite{Mosel1,Mosel2}.

\section{Vector mesons in medium}
The original suggestion by Hatsuda and Lee \cite{Hatsuda} that QCD
sum rules predict a significant lowering of the vector meson masses
in medium already at saturation density has created quite an
excitement at that time. We now understand that this prediction was
overly simplistic \cite{Leupold1,Leupold2} and that it has to be
replaced by the prediction that spectral strength of the vector
mesons moves down to lower invariant masses. The latter can be
achieved not only by a downwards mass-shift, but also by a
broadening, or a multi-humped structure, of the vector meson
spectral function. QCD sum rules thus do not predict specifics of
in-medium hadron spectral functions, but they do provide constraints
for model predictions. They also link -- although very indirectly --
the spectral functions seen in the hadronic world to the quark- and
gluon-condensates in the QCD world.

Thus, for a specific prediction of the in-medium spectral function a
hadronic model is necessary. It has been realized by now that in
particular resonance-hole contributions have a major influence on
hadronic in-medium properties. For the pion this has been known for
a long time; the Delta-hole model explains a large part of the
observed pion-nucleus interactions \cite{Ericsson-Weise}. As in this
particular case the most essential ingredient is the coupling
constant of the resonance-nucleon-meson vertex, which is obviously
large in the Delta-nucleon-pion case. Thus, at this point nucleon
resonance physics and in-medium physics meet!

For the $\rho$ meson the most essential resonance is the
$D_{13}(1520)$ excitation which -- according to the PDG -- has a
partial decay width into $\rho N$ of about 20\%. This directly
translates into a large coupling constant since this resonance lies
about 200 MeV below the nominal threshold for $\rho N$ decay
\emph{if} the decay width given is indeed correct\footnote{This
'subthreshold decay' has never explicitly been seen experimentally,
but has been inferred from an analysis of pion inelasticities. A
direct verification -- possible by a partial wave analysis of $2
\pi$ channels -- could help to clarify this essential point.}.

\begin{figure}[h]
\centerline{\includegraphics[width=20pc]{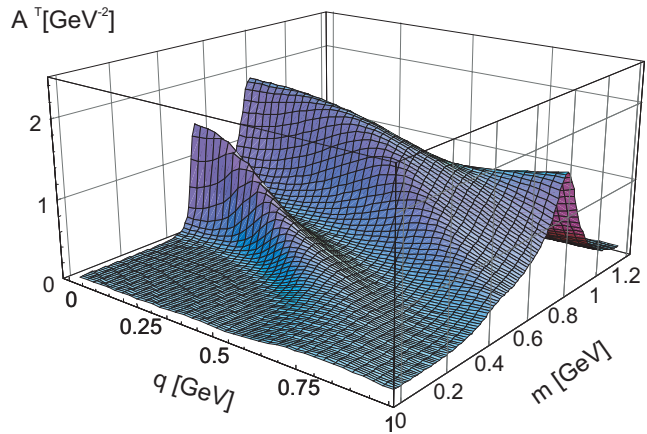}}
\caption{Selfconsistent (to all orders of density $\rho$) spectral
function of transverse $\rho$ mesons in nuclear matter at
equilibrium density $\rho_0$ (from \cite{Post}).} \label{rho-spect}
\end{figure}

The result of a calculation of the transverse part of the $\rho$
spectral function employing the 'official' coupling strength is
shown in Fig. \ref{rho-spect}.

The calculated spectral function shows significant structure that
directly reflects various nucleon resonances, the most dominant one
being indeed the $N(1520)$ resonance which leads to the sharp second
peak at invariant masses around 1500 MeV. One also notices that
there is a strong momentum-dependence in the spectral function and
that in particular this low-mass hump disappears at higher momenta
where the spectral function still shows considerable broadening
compared to the free case, but no longer the mass-shift seen at
small momenta \cite{Post1}.

\subsection{Omega Meson}
Exactly the same model has been applied recently to the in-medium
selfenergy of the $\omega$ meson. Early models for this meson had
mainly invoked a change of the pion cloud in medium; such studies
were restricted to tree-level diagrams with very limited basis space
(no resonances to dress the mesons involved) \cite{Klingl}.

A complication when applying the resonance-hole model also to the
$\omega$ meson lies in the fact that here the experimental knowledge
on the resonance-nucleon-omega coupling is still much more uncertain
than for the $\rho$ meson. Until a few years ago, no resonances were
known that decay into $N\omega$ and only recently experiments
\cite{Elsa-om} and analyses \cite{Penner,Lutz,Shklyar} have become
available. The results on the coupling still change from analysis to
analysis; it is clear, however, that background terms play a large
role here in addition to resonance couplings.

\begin{figure}[bp]
\centerline{\includegraphics[width=20pc]{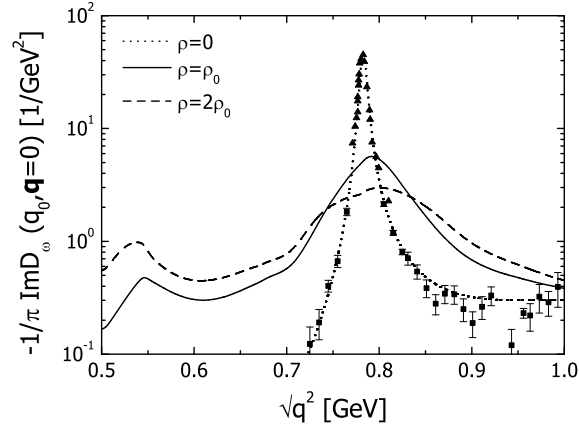}}
\caption{The spectral function for an $\omega$ meson at rest. The
appropriately normalized data points correspond to the reaction
$e^+e^- \to \omega \to 3 \pi$ in vacuum. Shown are results for
densities $\rho = 0$, $\rho= \rho_0 = 0.16$ fm$^3$ (solid) and $\rho
= 2 \rho_0$ (dashed)(from \cite{Muehlich}).} \label{om-spect}
\end{figure}

The results of a very recent analysis are shown in Fig.
\ref{om-spect}.
The calculations use - in lowest order in the density - the results
of a simultaneous coupled channel analysis of $(\gamma N)$ and $(\pi
N)$ reactions \cite{Shklyar} and in this way goes far beyond earlier
analyses which either contained no resonance-coupling at all
\cite{Klingl} or invoked VMD to relate the photon-nucleon coupling
to the $\omega N$ one \cite{Lutz,Postom}. The $\omega$ spectral
function at normal nuclear density is considerably broadened, but
hardly shows a shift at all (see Fig. \ref{om-spect}). This is due
to the comparatively small coupling to resonance-hole states that,
in addition, leads to a visible $N^*$hole component ($S_{11}(1535)$)
only far below the free $\omega$ mass.

As for the $\rho$ meson the transverse and the longitudinal
components behave quite differently in medium. Whereas the
longitudinal component changes its shape only very little the
transverse component gets significantly suppressed and broadened.
\begin{figure}[phtb]
\centerline{\includegraphics[width=20pc]{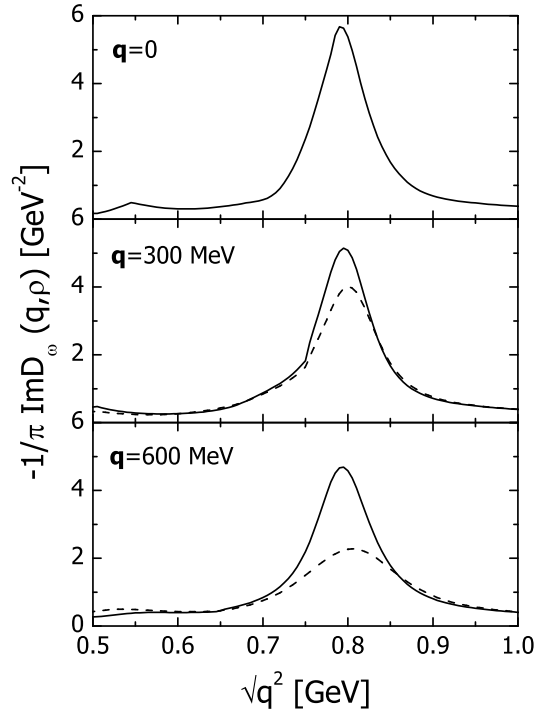}} \caption{The
longitudinal (solid) and transverse (dashed) spectral functions for
an $\omega$ meson at various momenta given in the figure (from
\cite{Muehlich}.} \label{omlt-spect}
\end{figure}

Recently, the CBELSA/TAPS experiment has reported evidence for a
change of the omega-in-medium spectral properties and has found a
lowering of the omega signal \cite{Trnka}. BUU calculations that
take the rescattering of the pion in the outgoing channel into
account can reproduce this signal when a lowering of the $\omega$
mass by about 14\% at $\rho_0$ is put in by hand \cite{Muehlichom1}.
The width obtained in the same experiment of about 55 MeV for slow
$\omega$s is of the same order as that predicted theoretically \cite
{Muehlich}, but is still determined by experimental resolution. The
observed downward shift of strength is contrary to what the
calculation mentioned above predicts. However, the experimental
result involves a product of spectral function and partial decay
width into the $\pi^0 \gamma$ channel that was used in the
experiment and it is this product that shows the downward trend. How
the spectral function by itself behaves in medium must be the
subject of further studies.

\section{Neutrino Reactions}

\begin{figure}[htb]
\centerline{\includegraphics[width=35pc]{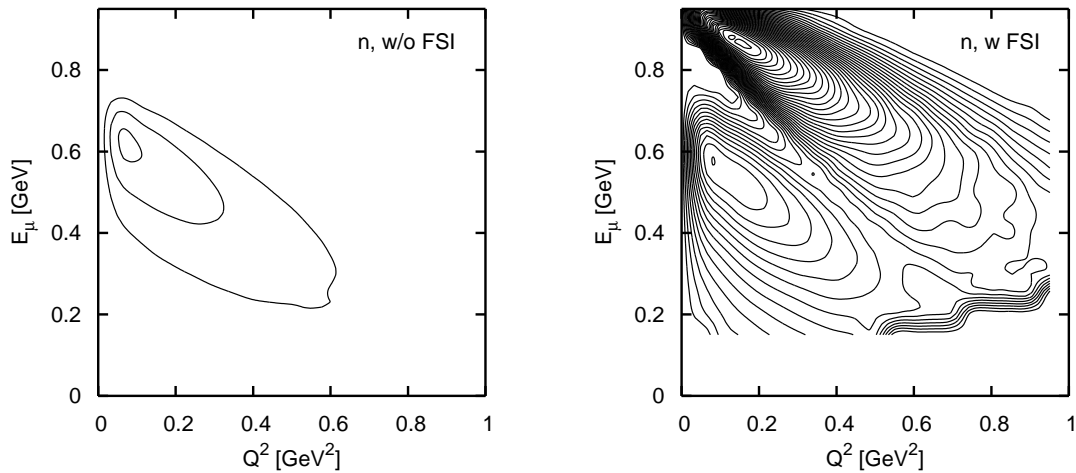}}
\caption{Double differential cross section $d\sigma/dQ^2\,dE_\mu$
for neutron knockout on $^{56}Fe$ at $E_\nu = 1$ GeV. Left: without
FSI, right: with FSI (from \cite{Leitner-CC}.} \label{n-knockout}
\end{figure}

A very dramatic example for the action of final state interactions
is provided by the charged current neutrino-induced neutron knockout
off nuclei. Since charged current (CC) interactions by themselves
always change the charge of the hit nucleon by one unit there cannot
be any CC knock-out neutrons in a quasielastic process. This is
indeed born out in the results of calculations (see Fig.
\ref{n-knockout}, left) \cite{Leitner-CC}. The few events visible in
that picture at $Q^2 \approx 0.05$ GeV$^2$ and $E_\mu \approx 0.6$
GeV stem from events where first a $\Delta^+$ is produced that then
decays into $\pi^+ n$.

Final state interactions (FSI) are now being implemented by means of
the GiBUU transport method \cite{GiBUU}. When these final state
interactions, that involve elastic and inelastic rescattering,
particle production and sidefeeding, are turned on this picture
changes dramatically (see Fig. \ref{n-knockout}, right). Now a
significant neutron knockout signal appears at $E_\mu \approx 0.9$
GeV with a long ridge in $Q^2$. In addition the $\Delta$-like events
now show also considerably more strength. The former effect is
caused by charge-transfer reactions where in a first interaction a
proton is knocked on that then travels through the nucleus and
transmits its energy and momentum to a hit neutron that is being
knocked out of the nucleus. The same applies to the $\Delta$-like
events: due to charge-exchange FSI now also the initial decay
channels $\Delta^+ \to \pi^0 p$ and $\Delta^{++} \to \pi^+ p$ can
contribute to final neutrons being knocked out.

\begin{figure}[htb]
\centerline{\includegraphics[width=10cm]{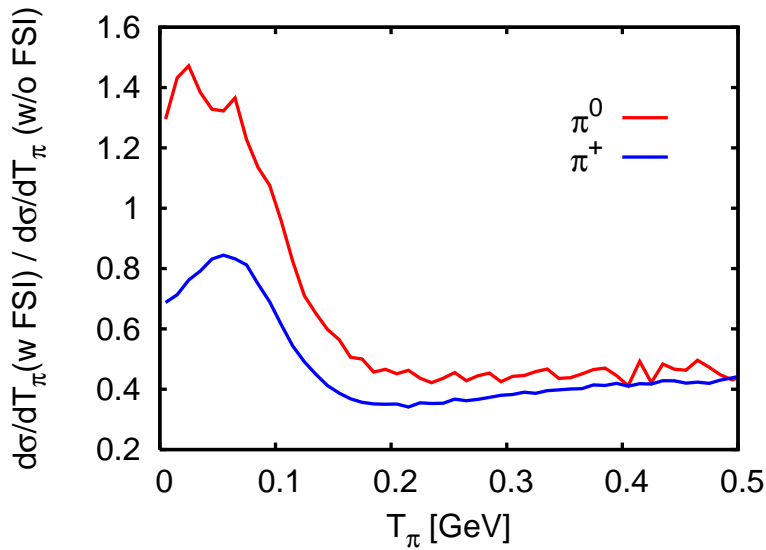}}
\caption{Ratio of energy-differential cross sections for pion
production with and without final state interactions for $\pi^+$
(lower curve) and $\pi^0$ (upper curve) for CC interactions of 1 GeV
neutrinos on $Fe$.} \label{Ratio}
\end{figure}

Details of the $\pi N$ interactions can be seen in Fig. \ref{Ratio}
that shows the ratios of the energy-differential pion production
cross sections with and without FSI. Besides an overall suppression
to a level of $R \approx 0.4$ for kinetic energies beyond about 200
MeV caused by pion absorption through the $\Delta$ resonance the
ratio shows a strong increase towards smaller energies that reflects
the slowing down of pions through rescattering. The drop at very
small energies is then caused by, at this low energy increasingly
important, many-body pion absorption. Particularly interesting is
that at low energies the $\pi^0$ curve rises significantly higher
than the $\pi^+$ ratio and even goes well above 1. This is due to a
coupled channel effect in the final state which always tends to go
from the stronger channel to the weaker one and in this case
converts positively charged pions into neutral ones.


\section{Hadron Formation}
A final, interesting example for in-medium effects shows up in the
high-energy regime and can be represented as the question how long
it takes to produce a particle for example in an electromagnetic
interaction process on the nucleon. In the lower-energy regime (a
few hundred MeV incoming energy) this time is often determined by
the lifetime of nucleon resonances. For example, pion production on
the nucleon predominantly proceeds through the $\Delta$ resonance
and the pion's formation time is thus given by the lifetime of the
$\Delta$. At higher energies, or invariant masses of the
nucleon-photon incoming system above about 2 GeV, specific nucleon
resonances no longer exist and these highly excited nucleon states
are described by string excitations. Phenomenological event
generators, that try to incorporate many processes of pQCD, such as
FRITIOF or PYTHIA, take care of the decay of such states and also
generate the space-time coordinates of such decays \cite{G-times}.

Transport methods have successfully been used to analyze such
high-energy hadron production experiments on nuclear targets
\cite{Falter}. In a simplified version of this model the prehadrons,
just after production through string breaking, experience no
interactions within a formation time, $\tau_f$, and after that the
full hadronic interactions. Since the latter take place only as long
as the hadron is still inside the nucleus the nuclear dimension
influences the interaction rate. The overall attenuation is then
sensitive to the number of collisions inside the medium weighted
with the formation time. The energy-loss is then, quite generally,
given by
\begin{equation}  \label{DeltaE}
\Delta E \sim \frac{L}{\lambda} \cdot F(L)
\end{equation}
with $L$ being the distance traveled in the nuclear medium and
$\lambda$ the prehadrons mean free path. The first factor is then
just the number of collisions. The function $F(L)$ takes into
account that an interaction can take place only if the particle has
already been formed
\begin{equation}
F(L) = \left\{ \begin{array}{c} \frac{L}{\tau_f} \quad { L/\tau_f < 1}\\
0 \quad {\rm otherwise}
\end{array} \right.
\end{equation}
In this simplified model the $L$ ($\sim A^{1/3}$) dependence is thus
already complicated. Indeed, an expansion of (\ref{DeltaE}) in terms
of powers of $A^{1/3}$ shows a mixture of various powers. The power
alone thus does not allow one to distinguish between different
models for the energy loss or absorption \cite{Falter,Accadi}. This
simple model can actually describe the data with a properly adjusted
formation time of about 0.5 fm/c in the restframe of the hadron
being formed, if in addition a reduced, but constant prehadronic
interaction cross section for leading hadrons is introduced.
However, it has been a problem to describe both the HERMES
experiments and the EMC attenuation data, taken at considerably
higher energies, with the same parameters \cite{Falter-thesis}.

Perturbative QCD predicts that the prehadronic cross sections should
rise linearly with time \cite{Dokshitzer}. We have recently
incorporated such a behavior into our calculations by using for the
prehadronic cross sections the expression \cite{Farrar}
\begin{equation}
\frac{\sigma_{\rm pre}}{\sigma_H} = \frac{n_q}{Q^2} \left(1 -
\frac{t - t_p}{t_f - t_p} \right) + \frac{t - t_p}{t_f - t_p}~.
\end{equation}
Here $t_p$ and $t_f$ denote the production time, i.e.\ the time at
which the quarks inside the hadron are produced through string
breaking, and the formation time, respectively. The quantity $n_q$
denotes the number of leading quarks, i.e. quarks present in the
original hit hadron.
\begin{figure}[hpbt]
\centerline{\includegraphics[width=20pc]{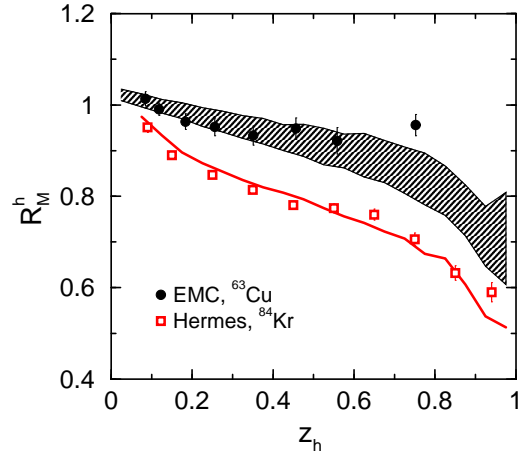}}
\caption{Nuclear attenuation ratio for the HERMES and the EMC
experiment. The shaded band for the EMC experiment gives the
uncertainty connected with averaging over the beam energies of 100
and 280 GeV, in line with the experimental analysis. (from
\cite{Gall-Ms})} \label{HERMES-EMC}
\end{figure}

Fig.\ \ref{HERMES-EMC} shows the attenuation as a function of $z =
E_{\rm hadron}/\nu$ demonstrates that such a model indeed gives a
very good description of the total hadron attenuation in both the
HERMES and the EMC experiments.

\section{Summary}

In this talk various aspects of in-medium effects have been
demonstrated. First, the importance of resonance-hole excitations
for calculations of in-medium hadronic spectral functions of the
$\rho$ and $\omega$ meson has been demonstrated. Since any in-medium
signal that involves hadrons in the final states is subject to final
state interactions these latter have to be treated with the same
accuracy as the in-medium effects themselves. This can nowadays be
achieved with semiclassical transport methods; usable
quantum-mechanical approaches do not exist for the description of
inclusive events. Using an example from neutrino-nucleus
interactions, charged current neutron knock-out, the overwhelming
influcence of final state interactions has been demonstrated.
Finally, the potential to determine hadronic formation times in
high-energy collisions of photons with nuclei has been pointed out.

\end{document}